\documentclass[twocolumn,showpacs,preprintnumbers,amsmath,amssymb,superscriptaddress,10pt,aps,prb]{revtex4-1}
\usepackage{epsfig,amsopn}
\usepackage{graphicx}
\usepackage{epstopdf}
\usepackage{amsmath,amssymb}
\usepackage{amsthm}
\usepackage{enumerate}
\usepackage{xcolor}

\begin{document}

\title{RKKY coupling in Weyl semimetal thin films}
\author{Sonu Verma}
\affiliation{Department of Physics, Indian Institute of Technology Kanpur, Kanpur 208016, India}
\author{Debasmita Giri}
\affiliation{Department of Physics, Indian Institute of Technology Kanpur, Kanpur 208016, India}
\author{H.A. Fertig}
\affiliation{Department of Physics, Indiana University Bloomington, Bloomington 47405, USA}
\author{Arijit Kundu}
\affiliation{Department of Physics, Indian Institute of Technology Kanpur, Kanpur 208016, India}

\begin{abstract}
We consider the effective coupling between impurity spins on surfaces of a thin-film Weyl semimetal within Ruderman-Kittel-Kasuya-Yoshida (RKKY) theory. If the spins are on the same surface, their coupling reflects the anisotropy and the spin-momentum locking of the Fermi arcs. By contrast when the spins are on opposite surfaces, their coupling is mediated by the Fermi arcs as well as by bulk states.  In this case the coupling is both surprisingly strong and strongly thickness dependent, with a maximum at an optimum thickness. We demonstrate our results using analytical solutions of states in the thin-film geometry, as well using a two-surface recursive Green's function analysis of the tight-binding model.
\end{abstract}

\maketitle

\section{Introduction}\label{sec:intro}
Weyl semimetals (WSMs) are three dimensional topological systems that host an even number of band-touching points (termed as \textit{Weyl nodes}) in the bulk spectrum, near which the low-energy excitations follow the relativistic Weyl equation~\cite{WSMreview}. Such Weyl quasiparticles have definite chirality and the chirality of these quasiparticles are given by the nature of the Weyl nodes, which can act as either sources or sinks of Berry curvature in the Brillouin zone. In a finite geometry, WSMs also host unique surface states, named as \textit{Fermi arc} states, whose projected Fermi surfaces are open arcs on each of the surfaces. Numerous materials have been predicted to be suitable candidates for WSMs , and a variety of experiments demonstrate their novel character.

Correlations functions impact many properties of these systems, and are of special interest because of
the unique helical nature of low-energy excitations in WSMs. Within this class of properties,
the effective interaction between two localized impurity spins introduced in such a system, mediated by the WSM conduction electrons, is described by the Ruderman-Kittel-Kasuya-Yoshida (RKKY) theory~\cite{RKKY}, and is  directly related to the spin-spin correlation function of  electrons within the system. Among solid-state materials, spin-orbit coupled systems ~\cite{RKKYSO1,RKKYSO2}, particularly topological ones, are particularly interesting in the way they mediate long-range -- and sometimes controllable -- coupling\cite{RKKYQSH1,RKKYQSH2,RKKYElectrical1,RKKY2DTI,RKKY2DTI2,RKKYTI,Reja_2017,Reja_2019,SonuRKKY} among spins. Work on RKKY interactions through bulk Weyl fermions~\cite{RKKYWSM1,RKKYWSM2,RKKYWSM3} show the interactions can be anisotropic and are in some circumstances weak, but generally carry signatures of the chiral nodes.

Significant attention has also been given to RKKY interactions on the surface of WSMs~\cite{RKKYWSMSur1,RKKYWSMSur2,RKKYWSMSur3}.
Surface states, at the Fermi energy of a WSM in a slab geometry, typically reside
at wavevectors which form arcs in the surface Brillouin zone.  These arcs join one Weyl node to the other, and typically disperse energetically perpendicular to a given arc, with different signs of the dispersion for each of the two physical surfaces. The essentially one dimensional character of the surface states results in a strong, highly anisotropic spin-spin correlation function, with similarly anisotropic RKKY interactions among spin
impurities adsorbed on the surface of a WSM.  Previous studies
have been largely confined to semi-infinite geometries, for which coupling {\it between}
surfaces cannot be modeled.  Such coupling is potentially significant,
as the Fermi arc states can be relatively weakly localized at their surfaces; moreover,
the penetration length of a Fermi arc state diverges as the surface wavevector
approaches the projection of a Weyl node. This can induce interesting physics due to
non-negligible coupling between spins on opposite surfaces of the WSM.

In this paper, we investigate the effects of such intersurface coupling by analyzing a slab geometry of {finite thickness}. Specifically, we examine effective spin-spin interactions due to the RKKY mechanism for two different situations. Firstly, when two spins are on the {\it same} surface of the WSM, surface electrons on the opposite surface can participate in their coupling. The resulting RKKY interaction reflects the anisotropy of the Fermi surface, and in the thick slab limit recovers previous results in which only a single surface was modeled~\cite{RKKYWSMSur1}. Secondly, when the spin impurities are on {\it opposite} surfaces of the slab, the resulting coupling depends strongly on the overlap of the Fermi arc states. We find that the resulting coupling is a non-monotonic function of the slab thickness, and a thin-film limit can be defined by observing when the coupling between the spins is strongest. In order to compute the coupling, we have developed a recursive Green's function scheme in which the elements of a Green's function on the surfaces can be computed essentially exactly with relatively high numerical efficiency. We show that analytical solutions for the WSM wavefunctions in a slab geometry verify the numerical results, and offer some insight into their qualitative behavior.

This manuscript is organized as follows. In Section II, we introduce the simple WSM model  used for our work and find analytical solutions for wavefunctions in a slab geometry with appropriate boundary conditions. In Section III, we briefly discuss the formal expression for RKKY interactions and our numerical scheme for computing them in a slab geometry of a tight-binding model.  Our numerical results are presented in Section IV, along with a comparison with analytical results. Finally we conclude with a summary and discussion
in Section V.

\section{Weyl Semimetal - Thin Film}\label{sec:wsmtf}
\subsection{Model Hamiltonian}
A minimal model of a WSM has two Weyl nodes at the Fermi energy and breaks time-reversal (TR) symmetry. For such a model, the low-energy Hamiltonian can be written using a two-band model. If the two bands represent spin states, then for a slab geometry, with the Weyl nodes separated along the momentum of one of the translational invariant (in-plane) directions,
%
the surface states (Fermi arcs) are  spin-polarized, resulting in completely spin-polarized surfaces of the slab. As the indirect spin-exchange interaction is only interesting when the ground-state is spin-{\it unpolarized}, the minimal model we consider must have at least two Fermi arcs on each surface, with the spin-polarizations of
each oriented such that the net spin density on either surface vanishes. If the two Fermi arcs on a surface have distinct locations in the surface Brillouin zone, then one has a total of four Weyl nodes in the bulk, each with a distinct location in momentum space. If the Fermi arcs join two {\it Dirac} nodes, then the Fermi
arcs will overlap in the surface Brillouin zone.  This latter situation was considered, for example, in Ref.~\cite{RKKYWSMSur1}.  In our work we confine our studies to the former case (i.e., Weyl semimetals).

\begin{figure}[t]
	\centering
	\includegraphics[width=0.4\textwidth]{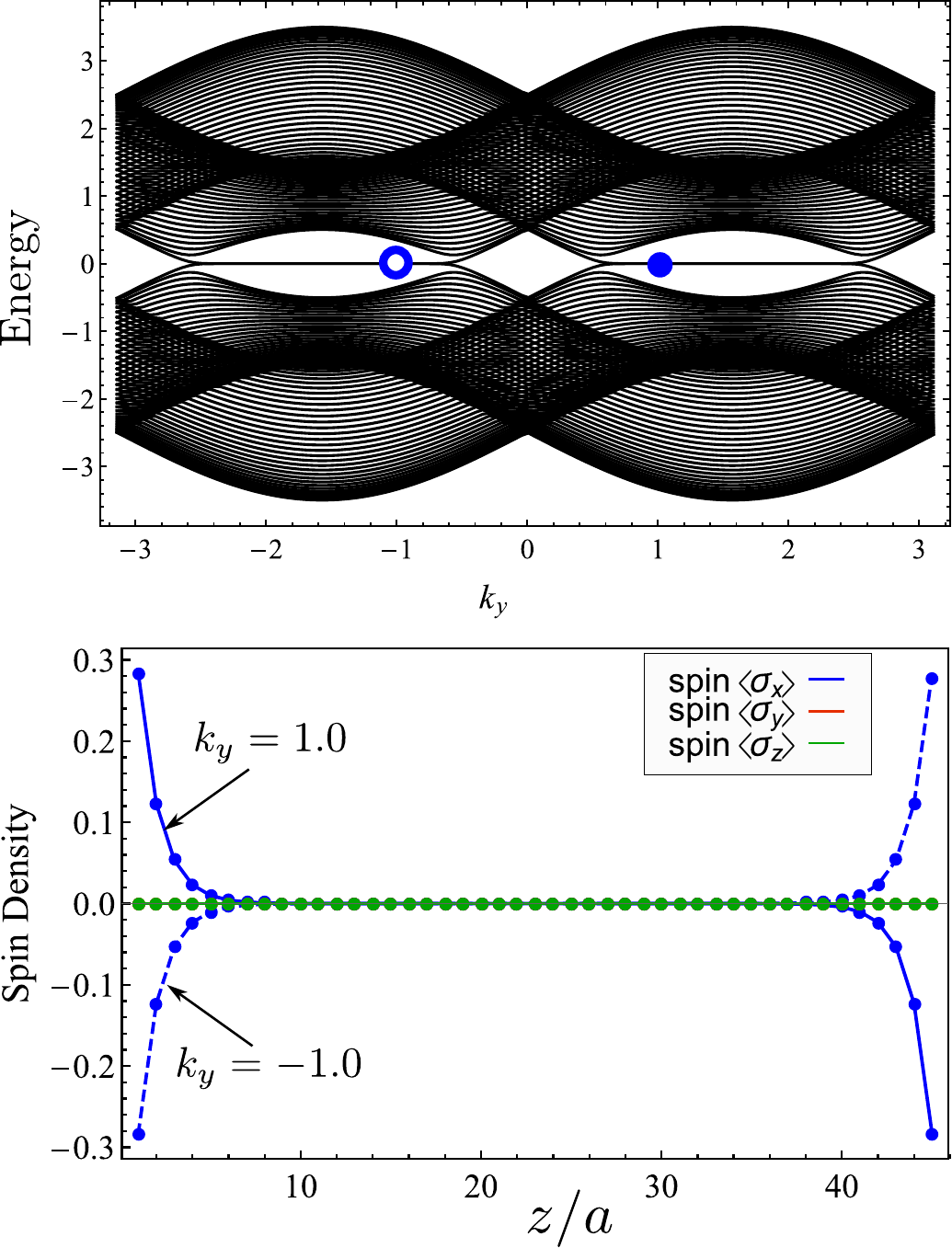}\\
	\caption{Top: For the WSM, Eq.~(\ref{eq:H1}), in a slab geometry with finite thickness in $z$ direction and for $k_x=0$, the band-structure (in the unit of $\lambda$) as a function of $k_y$ shows the four Weyl nodes in the bulk and the two Fermi arc joining them (the lattice spacing $a$ is taken to be unity). Bottom: The spin densities of the Fermi arc states are shown along various $z$ for two values of $k_y$ as pointed in the top figure. Parameters: $m=0.5~\lambda$, thickness $N_z=45$ lattice spacings.}\label{fig:bulk}
\end{figure}

Our starting point is a model Hamiltonian defined on a cubic lattice~\cite{WSMModel}.
The Hamiltonian preserves time-reversal symmetry (defined by the time reversal operator $ T= \textit{i} \sigma_y K $, with $ K $ the complex conjugation operator and $\sigma_y$  a Pauli matrix acting in the spin-space), but breaks inversion symmetry, and so has four degenerate Weyl nodes.  Specifically, we take
\begin{equation}\label{eq:H1}
H(k)=\lambda \sum_{\alpha =x,y,z} \sigma^\alpha \sin{k_\alpha} -\mu + \tau^y \sigma^y M_k.
\end{equation}
Here $M_k=m+2-\cos{k_x}-\cos{k_z} $, and $ \tau^\alpha $ are Pauli matrices acting in an orbital space. For $ | m | \leqslant \lambda $ the four Weyl nodes are located at $ \textbf{\textit{k}}=(0,\pm \pi/2 \pm k_0 ,0) $, where $k_0 = \pi/2-\sin^{-1}(m/\lambda)$.
On a given surface, the two Fermi arcs join the four Weyl nodes in a pairwise fashion, as illustrated in Fig.~(\ref{fig:bulk}). States of the two Fermi arcs are spin-polarized along $\sigma_x$ in opposite directions (i.e, they are eigenvectors of $\sigma_x$ with opposite eigenvalues for the two Fermi arcs). Furthermore, the two Fermi arcs, at low-energy, are dispersionless along the $k_y$ direction and have opposite velocities along the $x$ direction.

The Hamiltonian can be brought into a block diagonal form.  Writing
$H' = U H U^{\dagger}$, with the unitary matrix $U$ defined by
\begin{align}
U = \frac12\left(\begin{array}{cccc}
-1 & -i & -i & 1\\
1 & -i & i & 1\\
1 & i & -i & 1\\
-1 & i & i & 1
\end{array}\right),\label{eq:U}
\end{align}
one finds $H^{\prime}$ has two $2 \times 2$ blocks, where for each block (labeled by $\eta =\pm1$), the two-band Hamiltonian is
\begin{align}\label{eq:blokH}
H'_{\eta} &= \lambda(\sigma_y \sin k_x - \sigma_x \sin k_z) \nonumber \\
&+\eta \sigma_z (2 + m - \cos k_x - \cos k_z) - \lambda\sigma_z \sin k_y.
\end{align}
This is a particularly useful form, in which each block individually breaks time-reversal (TR) symmetry, while $T$ maps $H'_+$ to $H'_-$ (and vice-versa), so that the total Hamiltonian is TR symmetric. Each of the blocks has two Weyl nodes separated in momentum space, and on
a given surface they are joined by one Fermi arc. In principle, a system hosting many Fermi arcs on a surface should be structured in such a way that each joins two Weyl nodes; an
effective model of such a multi-Weyl node system could be written effectively as $H = H_1 \otimes H_2 \otimes \cdots$, where each of the blocks contains two Weyl nodes. 

To focus on the physics of the Fermi arcs, we expand the Hamiltonian to lowest non-trivial
order in $k_x$ and $k_z$, writing
$k_x \rightarrow q_x$ and $k_z \rightarrow q_z$.  Then from Eq.~(\ref{eq:blokH}) we obtain
\begin{align}\label{eq:newH}
H'_{\eta} &\approx \lambda(\sigma_y q_x - \sigma_x q_z) + \sigma_z M_{\eta}(k_y),
\end{align}
with $M_{\eta}(k_y) = m\eta - \lambda\sin k_y$. The four Weyl nodes are at
${\bf K}_{\eta,\xi} = (0,\eta\frac{\pi}{2}+\xi k_0,0)$ with $\eta, \xi =\pm 1$ and $k_0 = \cos^{-1}(m/\lambda)$.
For the $\eta=+1$ block, $M_+ < 0 $ between $k_y\in (\pi/2-k_0,\pi/2+k_0)$.
For a surface perpendicular to the $z$ direction, along the $k_y$ axis
these two points are connected by a Fermi arc.  For the $\eta=-1$ block, $M_- > 0 $ between $k_y \in (-\pi/2-k_0,-\pi/2+k_0)$, and again there is a Fermi arc connecting these points on the $k_y$ axis for the same surface. This situation is illustrated in Fig. \ref{fig:bulk}.
Note that for $H'$ (i.e., after the unitary transformation), states on the Fermi arcs are eigenvectors of $\sigma_y$ rather than $\sigma_x$.

\begin{figure}[t]
	\centering
	\includegraphics[width=0.45\textwidth]{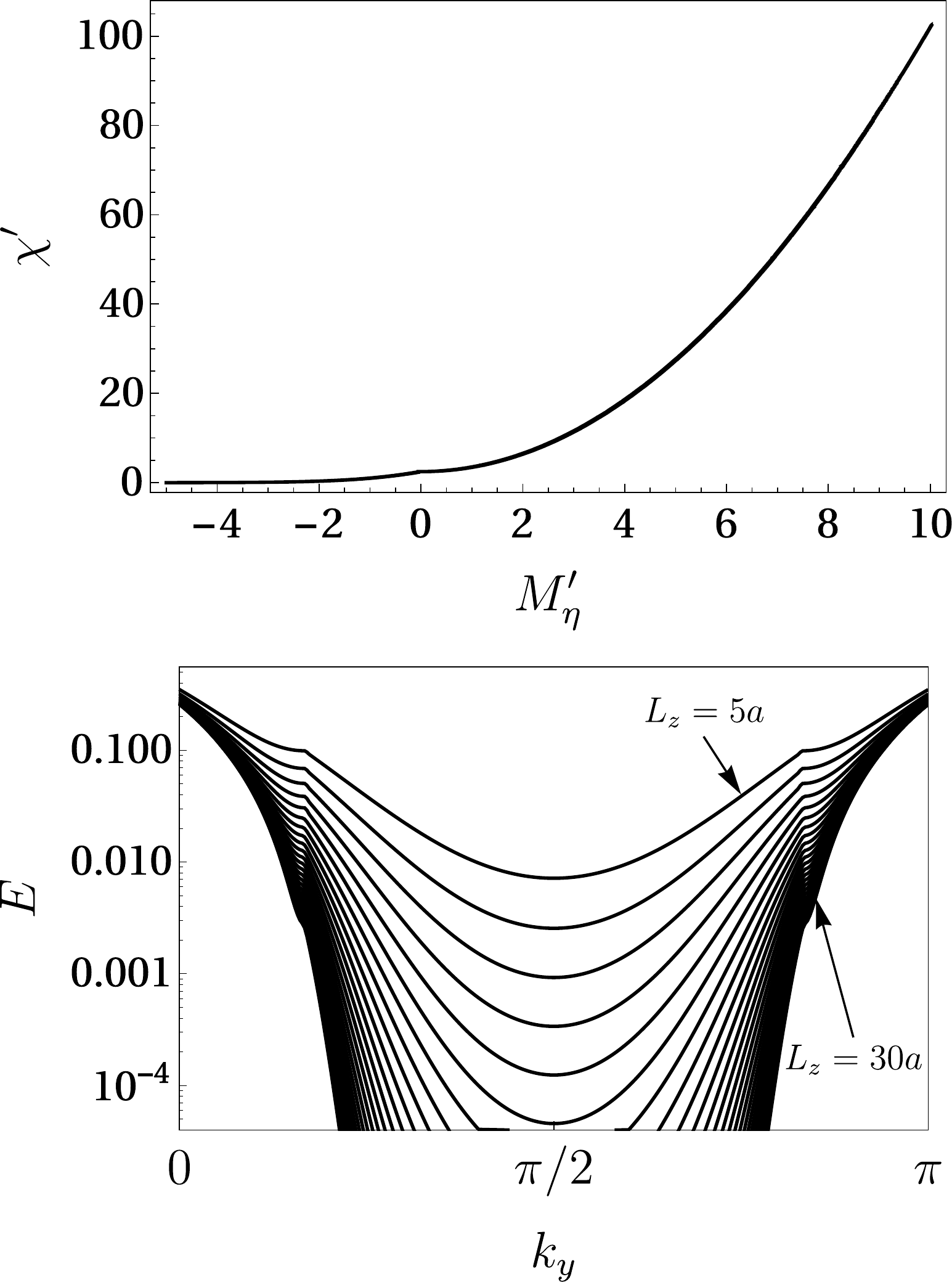}\\
	\caption{(Top) Minimum solution of $\chi'$ for values of $M^{\prime}_{\eta}$. When $M'_{\eta} \rightarrow -\infty$, $\chi' \rightarrow 0$, when $M'_{\eta} =0 $, $\chi' = (\pi/2)^2$ and for large $M'_{\eta}$, $\chi \approx M^{\prime 2}_{\eta} + \pi^2$. (Bottom) The lowest energy solution for $q_x=0$ for various values of $L_z$ from 5 to 30 are shown for half of the Brillouin zone, containing two Weyl nodes. The energy values decreases exponentially with increasing $L_z$ depicting surface states for $k_y$ between the Weyl nodes.}\label{fig:solchi}
\end{figure}

Near the Weyl nodes, if we can write the low-energy Hamiltonian in the form of $H=k_{\mu}A_{\mu\nu}\sigma_{\nu}$, then the chirality of the node is given by sgn(Det[$\mathbf{A}$]). Writing, $\textbf{\textit{k}} = (q_x,\eta\frac{\pi}{2}+\xi k_0 +q_y,q_z) $ and expanding to first order in $q_i$, we arrive at the low-energy Hamiltonian
\begin{align}
H^{{\rm low}}_{\eta\xi} &\approx \lambda(\sigma_y q_x - \sigma_x q_z) + \eta\xi  \alpha \sigma_z q_y,
\end{align}
with $\alpha =\sqrt{1-(m/\lambda)^2}$.  The chiralities of the four nodes may
then be written as ${\rm sgn}({\rm Det}[\mathbf{A}_{\eta,\sigma}])= -\eta\xi$.

\subsection{Infinite mass boundary condition}
To make progress analytically, we need to construct appropriate boundary conditions of the Dirac Hamiltonian Eq.~(\ref{eq:blokH}) for a slab geometry, such that the properties of the Fermi arc can be recovered. In general boundary conditions for Dirac equation can be cumbersome [REF], but our goal is to recover the properties of the surface modes (i.e, Fermi arc states). We construct boundary conditions by taking the Hamiltonian of the vacuum
(outside the slab, which extends from $z=0$ to $z=L_z$), similar to Eq.~(\ref{eq:newH}), except for the mass term, whose form is taken as $M^{\rm vac}_{\eta} = \eta m_0$, with $m_0 \rightarrow \infty$. This construction is required to ensure that for momentum between the Weyl nodes the effective mass term ($M_{\eta}(k_y)$) for the Weyl semimetal and the vacuum ($M_{\eta}^{\rm vac}$) are oppositely signed.  

The eigenfunctions for the Hamiltonian $H_{\rm vac} = \lambda(q_x\sigma_y - q_z\sigma_x) + M_{\eta}^{{\rm vac}}\sigma_z$ are
\begin{align}
\psi_{\rm vac} \propto \left(\begin{array}{c}
\lambda(q_z+i q_x) \\
M_{\eta}^{{\rm vac}}- E
\end{array} \right)e^{i(q_z z + q_x x)},
\end{align}
with eigenvalue $E =\pm \sqrt{m_0^2 +\lambda^2(q_z^2 + q_x^2)}$. For $m_0 \gg E$, the eigenfunctions are normalizeable if
\begin{align}
& q_z = i\kappa, \quad {\rm for} ~~ z\ge L_z, \nonumber\\
&q_z = -i\kappa, \quad {\rm for} ~~ z\le 0,\nonumber
\end{align}
with $\kappa = \sqrt{m_0^2 +q_x^2 -E^2}$. Thus, in the limit $m_0 \rightarrow \infty$, we have $\kappa \rightarrow m_0$. For $z>L_z$,
\begin{align}\label{eq:vacone}
\psi_{>} \propto \left(\begin{array}{c}
im_0+iq_x \\
\eta m_0 -E
\end{array} \right)e^{-m_0z} \approx \left(\begin{array}{c}
i\\
\eta
\end{array} \right)e^{-m_0z}.
\end{align}
For $z<0$,
\begin{align}\label{eq:vactwo}
\psi_{<} \propto \left(\begin{array}{c}
-im_0+iq_x \\
\eta m_0 -E
\end{array} \right)e^{m_0 z} \approx \left(\begin{array}{c}
i\\
-\eta
\end{array} \right)e^{m_0z}.
\end{align}
At $z=0, L_z$, these spinors become the Fermi arc wavefunctions, and are recognizable as eigenvectors of $\sigma_y$.

\begin{figure*}[ht]
	\centering
	\includegraphics[width=0.85\textwidth]{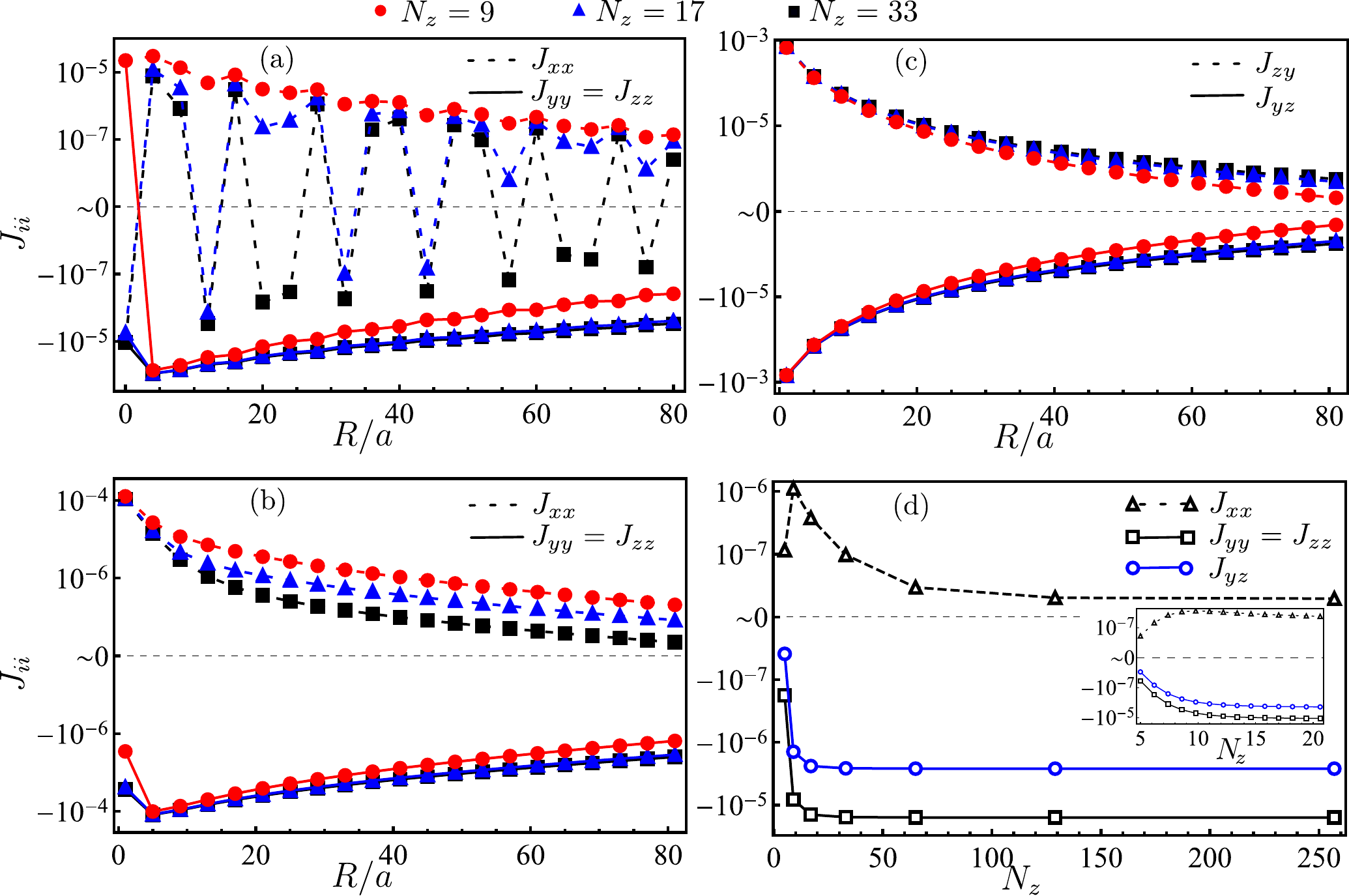}
	\caption{ The RKKY coupling between two spins (connected to the same orbital) put on the same surface (along $x$-direction \textit{i.e.} $\mathbf{R}=(R,0,0)$) of the WSM slab with (a) the analytical wave-functions and keeping only the $n=0$ bands, (b), (c) numerically evaluated Green's function in the real-space. With increasing thickness, the all components except $J_{xx}$ becomes essentially thickness independent after certain thickness, as shown in (d). Inset of (d) shows the RKKY coupling vs slab thickness calculated using analytical wave-functions and $n=0$ bands. Results shown are for $\mu=0$ (i.e., Fermi wavevector $k_F=0$)
and $m=0.5\lambda$.  $R/a=40$ in panel (d).
}\label{fig:top}
\end{figure*}

Matching the wavefunction $\psi(z)$ within the slab to these boundary forms yields the conditions
\begin{align}\label{bc}
\psi(z=0) \propto \psi_{{\rm}<}(z=0)~~{\rm and}~~\psi(z=L) \propto \psi_{{\rm}>}(z=L_z),
\end{align}
where
\begin{align}\label{eq:states}
\psi(z) =& a\left(\begin{array}{c}
\lambda(q_z+iq_x) \\
M_{\eta}(k_y) - E
\end{array} \right)e^{iq_z z}\nonumber\\
& + b\left(\begin{array}{c}
\lambda(-q_z+iq_x) \\
M_{\eta}(k_y) - E
\end{array} \right)e^{-iq_z z},
\end{align}
with $q_z =(1/\lambda)\sqrt{E^2 - M_{\eta}^2 -\lambda^2q_x^2}$. Non-trivial solutions of Eq.~\ref{bc} exists if

\begin{widetext}
\begin{align}
{\rm Det}\left(\begin{array}{cccc}
i & \lambda(q_z+iq_x) & \lambda(-q_z+iq_x)& 0\\
-\eta & M_{\eta}(k_y) - E & M_{\eta}(k_y) - E & 0\\
0 & \lambda(q_z+iq_x)e^{iq_z L_z} & \lambda(-q_z+iq_x)e^{-iq_z L_z}& i\\
0 & (M_{\eta}(k_y) - E)e^{iq_zL_z} & (M_{\eta}(k_y) - E)e^{-iq_z L_z} & \eta
\end{array} \right) = 0.
\end{align}
\end{widetext}
Simplifying this condition, we obtain a transcendental equation,
\begin{align}
&\frac{{\rm tanh}\left(L_z\sqrt{(M_{\eta}/\lambda)^2 - \chi}\right)}{L_z\sqrt{(M_{\eta}/\lambda)^2 -\chi}} = -\frac{\lambda}{L_z\eta M_{\eta}},
\label{eq:chi}
\end{align}
where $\chi = (E/\lambda)^2 - q_x^2$. For all real solutions $\chi$ of this equation, the energy has values $E = \pm\lambda\sqrt{\chi+q_x^2}$. No solutions of Eq. \ref{eq:chi} exist with $\chi<0$.

\begin{figure*}[ht]
	\centering
	\includegraphics[width=0.99\textwidth]{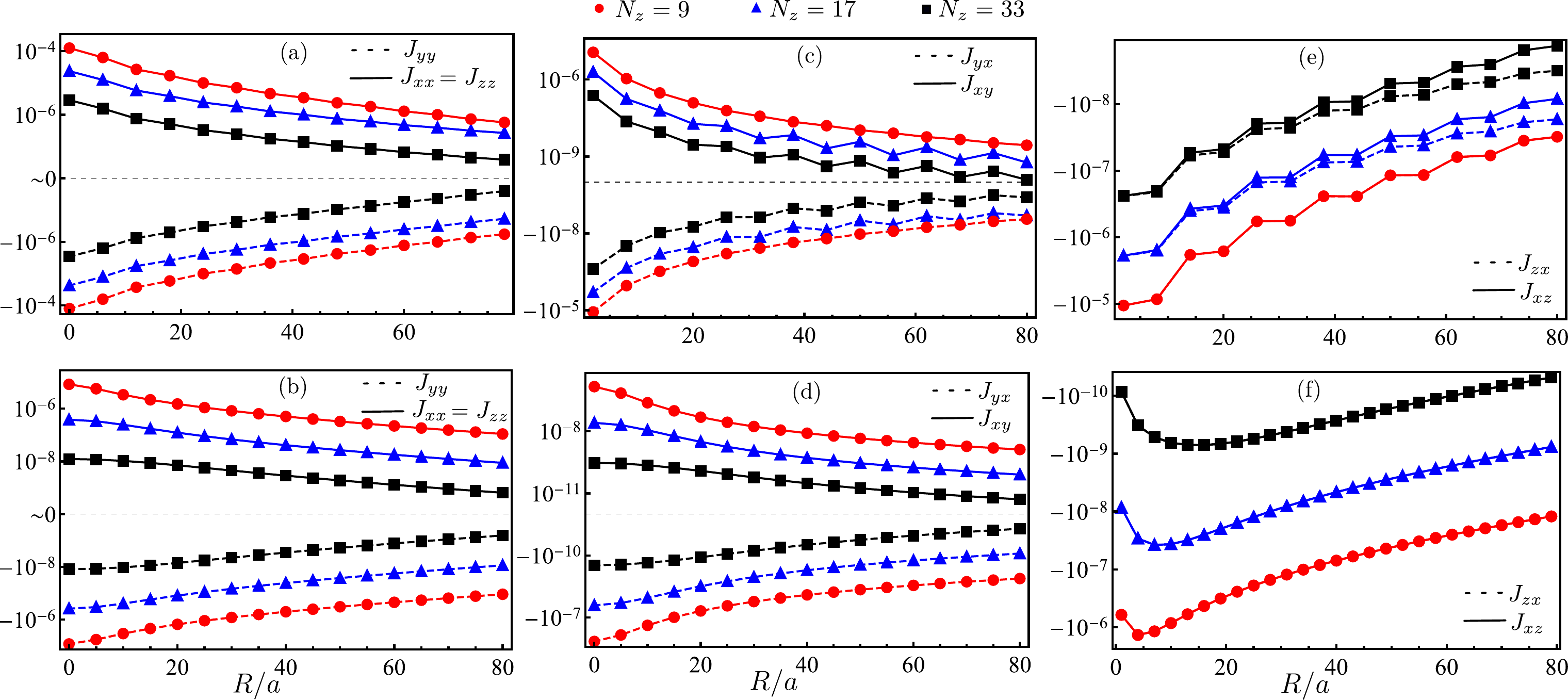}
	\caption{The RKKY coupling between two spins (connected to same orbital) on opposite surfaces of the WSM slab, with the positions of the two spins at $(x,y,z) = (0,0,0)$ and $ (R,0,L_z=N_za)$. (a), (c) and (e) are results for the analytical wavefunctions keeping only the $n=0$ bands. (b), (d) and (f) show numerically evaluated results from the Green's function approach. With increasing thickness, all components decrease rapidly (shown in more detail in Fig.~\ref{fig:withN} and tabulated in Table II). These results are for $\mu=0$
($k_F=0$). For all panels, $m=0.5~\lambda$.  
}\label{fig:two}
\end{figure*}

For bound-state solutions, i.e, when $q_z$ is imaginary, $\chi < (M_{\eta}/\lambda)^2$.  The left hand side of Eq. \ref{eq:chi} is a positive function with values between 0 and 1. Thus, such bound-state solutions are only possible when $\eta M_{\eta}<0$ as well as when $|L_zM_{\eta}/\lambda|>1$, i.e., when $|M_{\eta}|>\lambda/L_z$.
Defining $\chi' = L_z^2\chi$ and $M'_{\eta} = \eta L_zM_{\eta}/\lambda$, we rewrite
Eq. \ref{eq:chi} as
\begin{align}\label{eq:chim}
&\frac{{\rm tanh}(\sqrt{M_{\eta}^{\prime 2} - \chi'})}{\sqrt{M^{\prime 2}_{\eta} -\chi'}} = -\frac{1}{M^{\prime }_{\eta}}.
\end{align}
The various solutions of $\chi'$ from Eq. \ref{eq:chim} can be labeled by an index $n=0,1,..$ (with increasing values of $n$ corresponding to larger values of $\chi'$) and the corresponding energy solutions $E_{n,\pm}(q_x,q_y) = \pm \lambda \sqrt{\chi_n + q_x^2}$ gives rise to particle-hole symmetric bands. The minimum solution of $\chi'$ is shown in Fig.~\ref{fig:solchi}. The bands with $n=0$ contains all the Fermi arc states (when $k_y$ is between the Weyl nodes, in the Fermi arc interval) and low-energy bulk states (when $k_y$ is outside the interval).

The coefficient $(a/b)$ for the states, Eq.~(\ref{eq:states}), can be found from the boundary conditions at $z=0$ to be
\begin{align}
\frac{a}{b} = -\frac{M_{\eta} -E + \eta \lambda(q_x +i q_z)}{M_{\eta} -E + \eta \lambda(q_x - i q_z)}.
\end{align}
We can then write down the wavefunctions. Defining
$K = M_{\eta}(k_y) -E, f=\lambda(q_x-iq_z), g=\lambda(q_x+iq_z)$, one finds
\begin{widetext}
\begin{align}\label{eq:wf1}
|\psi \rangle &= \frac{1}{\sqrt{N}} \left\{ (K+\eta g)\left(\begin{array}{c}
i f \\
K
\end{array} \right)e^{iq_z z} +(K+\eta f)\left(\begin{array}{c}
-i g \\
-K
\end{array} \right)e^{-iq_z z} \right\}.
\end{align}
\end{widetext}
For real $q_z = \sqrt{\chi - M_{\eta}(k_y)^2}$ (when $\chi>m^2$, $f=g^*$) the normalization factor has the form
\begin{align}
N =& 2|K+\eta f|^2(K^2+|f|^2)L\nonumber \\
&+{\rm Im}\left[(K+\eta f)^2(K^2+g^2)\left(\frac{e^{-2iLq_z}-1}{q_z}\right) \right].
\end{align}
For purely imaginary $q_z=i\kappa$ (when $\chi<m^2$),$f=q_x+\kappa$, $g=q_x-\kappa$,
\begin{align}
N =&-2(K+\eta f)(K+\eta g)(K^2+g f) L\nonumber \\
&+[(K+\eta g)^2(f^2+K^2)e^{-\kappa L}\nonumber\\
&+(K+\eta f)^2(g^2+K^2) e^{\kappa L}]\frac{\sinh(\kappa L)}{\kappa}.
\end{align}
These are the full solutions of the low-energy states of the WSM slab in the rotated basis (Eq.~(\ref{eq:U})). Once written in the original basis, these solutions correctly reproduce the spin configuration of the Fermi arc states.

\section{RKKY interaction and Recursive Green's function}\label{sec:recgf}
Ruderman-Kittel-Kasuya-Yoshida (RKKY) theory~\cite{RKKY} describes the effective coupling between
two impurity spins ${\bf S}_1$ and ${\bf S}_2$ in a metal mediated by the conduction electrons.
The spins, located respectively at ${\bf r}_1$ and ${\bf r}_2$,
are typically treated as classical magnetic moments, and
are assumed to be coupled to the electrons by $sd$ Hamiltonians,
$H_{sd}=J{\bf S}_i \cdot {\bf s}({\bf r}_i)$ ($i=1,2$), where ${\bf s}({\bf r}_i)$ is
the conduction electron spin density at the location of impurity spin $i$.
For small $J$ the resulting impurity spin interaction becomes
\begin{align}
H_{\text{RKKY}} &= -\frac{J^2}{\pi} \int_{-\infty}^{E_F} d\omega \text{Tr}[(\mathbf{S}_1.\mathbf{\sigma})G(\mathbf{r}_{12};\omega + i0+)(\mathbf{S}_2.\mathbf{\sigma})\nonumber\\
&~~~~~~~~~~~~~~~~~~~~~~~~~~\times G(-\mathbf{r}_{12};\omega + i0+)]\label{eq:rkky}\\
& \equiv \sum_{i,j = x,y,z} J_{ij} S_{1i}S_{2j}\label{eq:corr},
\end{align}
where $\mathbf{r}_{12}$ is the separation of the two spins and
$G(\mathbf{r}_{12};\omega + i0+)$ is the real space Green's function for the unperturbed electron system. The resultant $J_{ij}$ is essentially the electronic spin-spin correlation matrix. In all of our results we show $J_{ij}$ in the unit of $J^2$.

Details of the particular electron system in which the impurity spins are embedded enter
the calculation through $G(\mathbf{r}_{12};\omega + i0+)$.
For our WSM system, we will proceed in two ways.
First, we will directly compute $G$ in momentum space from the low-energy Hamiltonian wavefunctions Eq.~(\ref{eq:wf1}), and then Fourier transform the expression
to obtain the needed real-space Green's function. Our second approach is more
numerical, and involves inverting the tight-binding model, Eq.~(\ref{eq:H1}).
In this approach the discrete translational invariance in the $x-y$ plane of the
slab geometry allows, for each two dimensional wavevector, independent computation of the Green's function.
A computation of the real space Green's function then follows from a Fourier transform.
For the results we present in the next section, we also restrict ourselves to considering impurities which are exchange-coupled to the same orbital of the two-orbital model, Eq.~\ref{eq:H1}, which captures the essential physics of interest. For the case of the semi-analytical model of the last section, for which the atomic-scale structure is not
included, we assume
the impurities to be exchange-coupled to the conduction electrons within a small region (of thickness of one lattice spacing) on each surface.

Before proceeding to our results, we use the remainder of this section to outline the
recursive Green's function method we use for our fully numerical studies.
We are interested in the coupling between impurities placed on the surfaces,
so that in the computation of $G(\mathbf{r}_{12};\omega + i0+)$ one only actually
needs the Green's function for sites ${\bf r}_1$ and ${\bf r}_2$ on the slab surfaces.
Following Ref.~\onlinecite{Keshav}, we can compute the two dimensional Fourier transform of this,
$G_{ij}(\omega,k_x,k_y)$, where $i$ and $j$ label the surfaces of the slab
on which ${\bf r}_1$ and ${\bf r}_2$ reside, respectively.

For a slab geometry of $N_z$ number of sites in the $z$ direction, we re-write the
tight-binding Hamiltonian (Eq. \ref{eq:H1}) in the form
\begin{align}\label{eq:slab}
H(\vec{k}_{||})=\sum_j &\left(\psi_{j}^{\dagger}(\vec{k}_{||})A(\vec{k}_{||})\psi_{j+1}(\vec{k}_{||}) + {\rm h.c.}\right.\nonumber \\
&\left.+ \psi_{j}^{\dagger}(\vec{k}_{||})h_{j,j}(\vec{k}_{||})\psi_{j}(\vec{k}_{||}) \right),
\end{align}
where $\vec{k}_{||}=(k_x,k_y)$, which are good quantum numbers. This allows us to write the Hamiltonian in the form of
\begin{align}
H &= \left(\begin{array}{ccccc}
h & A & 0 & .  & 0\\
A^{\dagger} & h & A &  . & 0\\
0 & A^{\dagger} & h &  . & 0\\
. & . & . & .  & .\\
0 & 0 & 0 &  . & h\\
\end{array}\right)
\end{align}
and the Green's function is evaluated from the equation
\begin{align}\label{eq:gdef}
\left(\omega\mathbb{I}-H(\vec{k}_{||})\right)G(\vec{k}_{||},\omega)=\mathbb{I}.
\end{align}
When $N_z = 1+2^k$, the above set of equations can be recast in the form
\begin{align}
\left( \omega\mathbb{I} -h' \right)G' = \mathbb{I},
\end{align}
with
\begin{align}
h' = \left(\begin{array}{cc}
h^{(k)} & A^{(k)} \\
A^{\dagger(k)} & h^{(k)} \\
\end{array}\right), \quad G' = \left(\begin{array}{cc}
G_{11} & G_{1N_z} \\
G_{N_z1} & G_{1N_z} \\
\end{array}\right),
\end{align}
where the $h_t^{(k)},h_b^{(k)}$ and $A^{(k)}$ are found by recursively solving
\begin{align}
A^{(i+1)} =A^{(i)}(\omega-h^{(0)})^{-1}A^{(i)}, \\
h^{(i+1)} =h^{(i)}+A^{\dagger(i)}(\omega-h^{(i)})A^{(i)}, \\
h_{t}^{(i+1)} = h_{t}^{(i)}+A^{(i)}(\omega-h^{(i)})^{-1}A^{\dagger(i)},\\
h_{b}^{(i+1)} = h_{b}^{(i)}+ A^{\dagger(i)}(\omega-h^{(i)})^{-1}A^{(i)},
\end{align}
with $h_t^{(0)}=h_{b}^{(0)}=h^{(0)}=h$ and $A^{(0)}=A$. This yields the two surface Green's functions $G(\vec{k}_{||})_{1,1}$ and $G(\vec{k}_{||})_{N_z,N_z}$ as well as their connections $G(\vec{k}_{||})_{1,N_z}$ and $G(\vec{k}_{||})_{N_z,1}$ without requiring
a solution for the full Green's function.

\begin{figure}[t]
	\centering
	\includegraphics[width=0.45\textwidth]{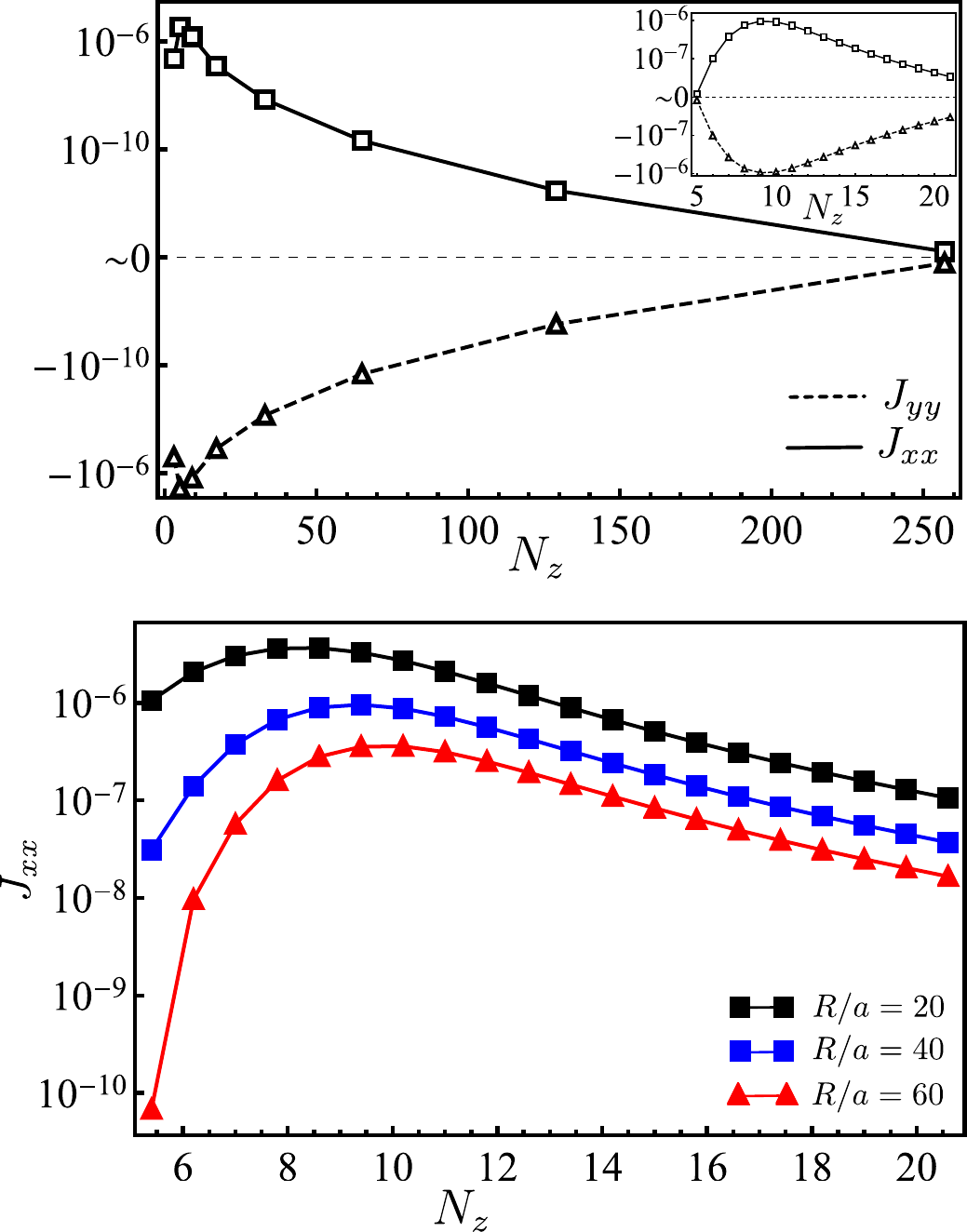}
	\caption{Top main panel: The RKKY coupling between two spins (connected to same orbital) on  opposite surfaces of the WSM slab of thickness $L_z=N_za$ and placed at (0,0,0) and $(R = 40a,0,L_z)$  as a function of $N_z$, evaluated using numerical Green's function method. Inset: Results from analytical wavefunctions, keeping only the $n=0$ bands. With increasing thickness, all components decrease rapidly (see discussion in the main text). Bottom panel: The RKKY coupling between two spins (connected to same orbital) on opposite surfaces of the WSM slab of thickness $N_za$ placed at three lateral distances $(R = 20a,0,L_z)$, $(R = 40a,0,L_z)$, $(R = 60a,0,L_z)$  as a function of $N_z$.  These results are computed using the analytical wavefunctions keeping only the $n=0$ bands. Note that the spin-spin couplings peak at slightly different slab thickness for different lateral separations between them. Other parameters are same as in Fig.~\ref{fig:two}.}\label{fig:withN}
\end{figure}

\begin{figure}[ht]
	\centering
	\includegraphics[width=0.45\textwidth]{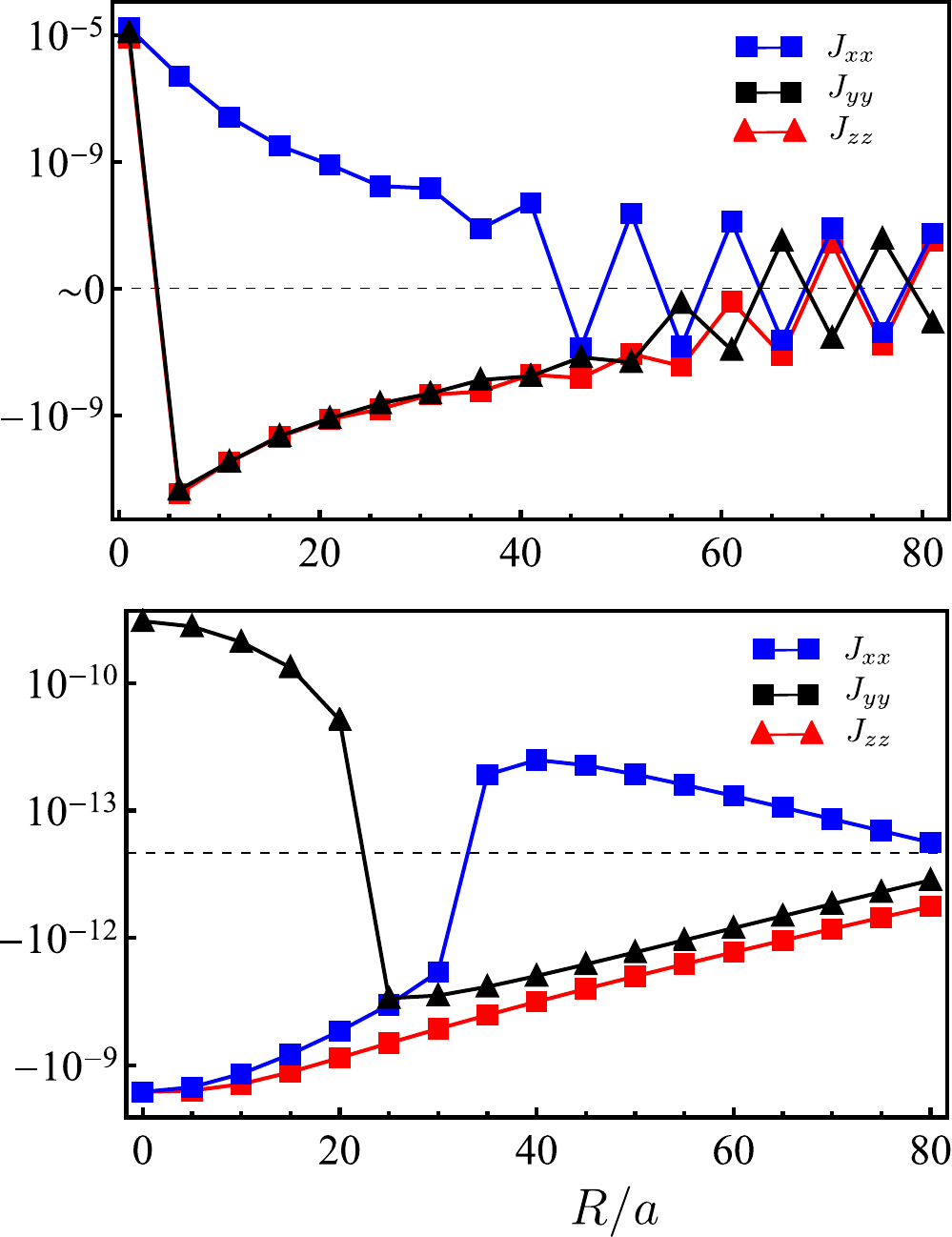}
	\caption{RKKY coupling for slab with surfaces perpendicular to the inter-Weyl node separation ($y$ direction in our model) for which there are no Fermi arcs.  $m=0.5\lambda$, $N_z=33$.
Top panel: Spins on the same surface.  Bottom panel: Spins on opposite surfaces. In comparison with Figs.~\ref{fig:top} and ~\ref{fig:two} the couplings are very small. For larger $R$, as the $J_{ii}$ become very small, the oscillations are possibly due to the numerical inaccuracy.
}\label{fig:xzsurface}
\end{figure}


\begin{figure*}[ht]
	\centering
	\includegraphics[width=0.95\textwidth]{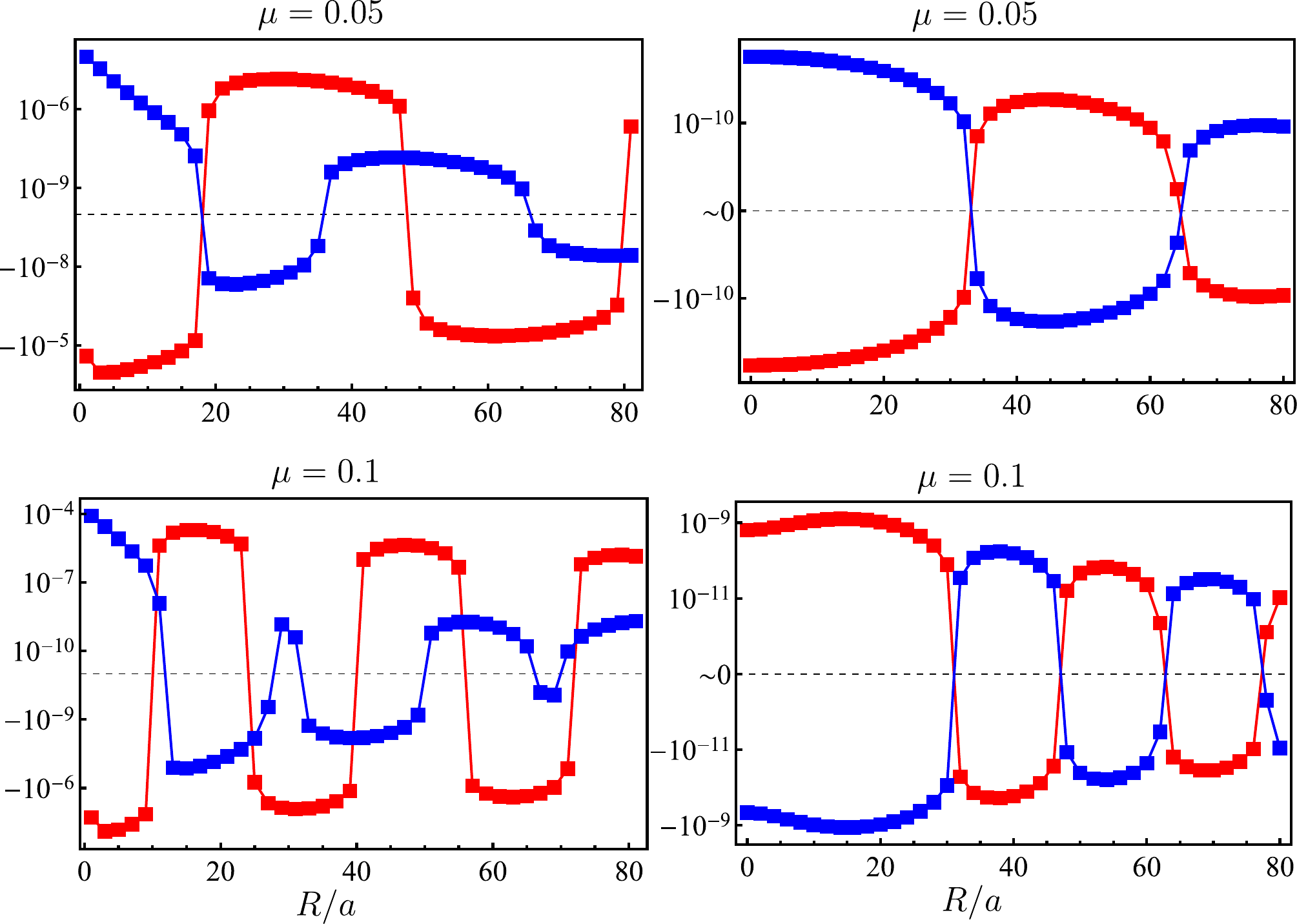}
	\caption{The diagonal elements of coupling matrix at a finite chemical potential (given by $\mu$ in units of $\lambda$), showing $2k_F$ oscillation. The left column shows the results for the spins on the same surface of the slab and the right column shows the results when the spins are on opposite surfaces. The slab thickness $N_z=33$ and all other parameters are same as in Figs.~\ref{fig:top} and ~\ref{fig:two}.}\label{fig:finitemu}
\end{figure*}

\section{Numerical results}\label{sec:num}
In this section, we present our results for the specific cases when ({\it i}) the magnetic impurities are on the same surface of the WSM, and ({\it ii}) when the magnetic impurities are on opposite surfaces of the WSM. In both cases we compute the RKKY interaction using the recursive Green's function method outlined in the last section, as well as using the analytical wavefunctions of the $n=0$ bands of Sec.~\ref{sec:wsmtf}, which contain the Fermi arc states, and compare the results.

\subsection{Impurities on a single surface}
As discussed above, the Fermi arc states disperse in energy along $\vec{k}$
perpendicular to the arc itself.
As a result these states have a highly asymmetric velocity, with $v_y\ll v_x$.
The effective interaction between spin impurities placed on the same surface reflects this
strong asymmetry. For a semi-infinite system, it can be shown for large $r$ that the elements of the Green's function $G(\vec{r})$ asymptotically vanish as $\sim 1/r^2$ when $\vec{r}$ is on the surface and
$\vec{r} \parallel \hat{y}$. By contrast, they fall off as $~\sim 1/r$ when $\vec{r}$
points along the $\hat{x}$ direction~\cite{RKKYWSMSur1}; the difference is a consequence of the (nearly)
unidirectional dispersion of the Fermi arc energies.
This results in the strongest RKKY coupling for impurities
separated along
the $\hat{x}$ direction, and in what follows we focus on separations along this direction.
Moreover,
for a thick enough sample (when the presence of the other surface may be neglected), states in each of the Fermi arcs are spin-polarized (along the direction of $\sigma_x$ in our model) and are chiral in their dispersion (i.e, the energy is proportional to $\pm k_x$ for $\eta=\pm1$). One expects in this case that the RKKY interaction $J_{xx}$ will be vanishingly small~\cite{RKKYQSH1,SonuRKKY,RKKYWSMSur1}.

These expectations may be understood as follows.
The Fermi arcs, for a thick sample, are exponentially confined to a surface at $z=0$ with an approximate wave function (see Eqs ~(\ref{eq:vacone}) and (\ref{eq:vactwo})):
\begin{align}
\psi_{{\rm FA}}(\textbf{r}) \approx e^{i k_x x} e^{i k_y y} e^{-M_{\eta}(k_y) \frac{z}{\lambda}}\left(\begin{array}{c}
i \\
\eta \\
\end{array}\right),
\end{align}
where $M_{\eta}(k_y) = m\eta -\lambda\sin(k_y)$.  These wavefunctions allow us to write an effective Green's function from the Fermi arc on this two-dimensional surface in the form
\begin{align}
G_{\eta}^{{\rm FA}}(\omega + i\delta;\textbf{r}) &= (\sigma_0-\eta\sigma_y)\int \frac{d^2 k}{(2\pi)^2}\frac{e^{i k_x x} e^{i k_y y}}{\omega -\eta v_F k_x +i\delta}\nonumber\\
&\times\frac{M_{\eta}(k_y)}{\lambda}\theta(k_y-k_{\eta,1})\theta(k_{\eta,2}-k_y),
\end{align}
where the Fermi arcs exists between $k_{\eta,1}= \eta\frac{\pi}{2} - k_0$ and $k_{\eta,2} = \eta\frac{\pi}{2} + k_0$. The $k_y$ momentum can be integrated between $k_{\eta,1}$ and $k_{\eta,2}$ and performing the $k_x$ integral one obtains,
\begin{align}\label{eq:GFFA}
G^{{\rm FA}}_{\eta} (\omega + i\delta; \textbf{\textit{r}}) = i\eta\frac{\eta \sigma_y - \sigma_0}{v_F^2} e^{i(\omega+i\delta) \frac{|x|}{v_F}}I(y)\theta(\eta x),
\end{align}
where, the Fermi velocities of the Weyl nodes $v_F = \lambda$. In the limit $\delta \rightarrow 0+$,
\begin{align}
&I(y) = e^{i\pi y/2}\Gamma(y),\\
&\text{with}~~ \Gamma(y)=\frac{yv_F\sin k_0 \cos (yk_0) - m\sin (yk_0)}{y(y^2-1)},
\end{align}
and \textbf{\textit{r}} = $(x,y)$. This approximate form for the Green's function is useful in determining the asymptotic behavior of the RKKY interaction, as we show briefly in the next section (see also Ref.~\onlinecite{RKKYWSMSur1}). Note that the step function in Eq.~(\ref{eq:GFFA}) implements the chiralities of the Fermi arcs. As $\eta=\pm$, in the full 4$\times$4 orbital/spin space, the Green's function is then
\begin{align}
G^{{\rm FA}}( \omega; \mathbf{r}) = \left(\begin{array}{cc}
G^{{\rm FA}}( \omega; \mathbf{r})_+ & 0\\
0 & G^{{\rm FA}}( \omega; \mathbf{r})_-
\end{array}\right).
\end{align}
With this expression
it is straightforward to work out the RKKY integral (Eq. \ref{eq:rkky})
and show that, in the original basis of the Hamiltonian, Eq.~(\ref{eq:H1}), the elements of the correlation matrix, Eq.~(\ref{eq:corr}), are given by
\begin{align}
& J_{xx}=0; ~~ J_{yy}=J_{zz} = \frac{J^2\Gamma (y)^2}{2\pi^3xv_F^4}\cos(2k_F x +\pi y),\\
& J_{zy}=-J_{yz} =  \frac{J^2\Gamma (y)^2}{2\pi^3xv_F^4}\sin(2k_F x +\pi y).
\label{eq:FAJs}
\end{align}
Other off-diagonal components vanish. In this limit, the separation of the \textit{pairs} of the Weyl nodes (taken as $\pi/a$ in our model) does not alter the result.

The symmetries apparent in Eqs. \ref{eq:FAJs} can be understood from the original tight-binding system as we discuss in Appendix A, and are captured by our results, as
illustrated in Fig.~\ref{fig:top}.
In particular
the value of $J_{xx}$ decreases with increasing thickness of the sample whereas the
other diagonal components become constant with increasing thickness. In contrast to the analytical result, $J_{zz}$ and $J_{yy}$ fall off asymptotically roughly as $1/x^2$, as listed in Table 1.  This difference from Eqs. \ref{eq:FAJs} is expected due to the curvature of the Fermi arcs~\cite{RKKYWSMSur1}.  Off-diagonal components other than $J_{zy}$ and $J_{yz}$ (not shown) are several orders of magnitude smaller than these quantities, as expected from the above
analysis. Note also the qualitative agreement between results from our semi-analytical model and the tight-binding computation.

\begin{center}
	\begin{table}[h]
		\begin{tabular}{ | c | c | c |}
			\hline
			& $N_z$ & $\alpha$  \\ \hline
			$J_{xx}$  & 9 & $\approx$ 2.53 \\
			\hline
			$J_{xx}$  & 17 & $\approx$ 2.29 \\
			\hline
			$J_{xx}$  & 33 & $\approx$ 2.26 \\
			\hline			
		\end{tabular}~~~~~~~
			\begin{tabular}{ | c | c | c |}
		\hline
		& $N_z$ & $\alpha$  \\
		 \hline
		$J_{zz}$  & 9 & $\approx$ 2.48 \\
		\hline
		$J_{zz}$  & 17 & $\approx$ 2.10 \\
		\hline
		$J_{zz}$  & 33 & $\approx$ 2.04 \\
		\hline
	\end{tabular}
	
		\caption{Assuming the couplings $J_{ii}$ between the impurity spins on the same surface go as $\sim R^{-\alpha}$, the best fitted value of $\alpha$ is quoted for the result of Fig.~\ref{fig:top}(b) for $R/a$ between 50 and 80.}
	\end{table}
\end{center}

The results for $J_{xx}$ obtained from the tight-binding computation are particularly interesting.  In contrast to the result for straight Fermi arcs in a single surface
system derived above, in the slab geometry $J_{xx}$
remains non-zero and falls off rather slowly (see Table 1) with the distance between the impurity spins.  (Similar behavior is found in our analytical model at small thickness.)  The discrepancy can be attributed to two possible effects: (i) interactions mediated by the bulk states which were not included in the simple Fermi arc analysis, and (ii) the presence of the second surface. Interestingly, Fig.~\ref{fig:top}(b) shows that $J_{xx}$ vanishes rapidly with increasing thickness, which clearly favors mechanism (ii). Fig.~\ref{fig:top}(d) also manifests a critical thickness $L_c$, defined by the width $L_z$ at which $J_{xx}$ attains its maximum value before decaying sharply with further increase. This critical thickness can be used to define a ``thin-film limit'' of the system, for which the effects of having two surfaces are maximal.  Noting that $k_0$ is the only relevant momentum scale, we expect the thin film limit to scale as $L_c \sim 1/k_0$. A numerical verification of this hypothesis is presented in Appendix B.

\subsection{Impurities on opposite surfaces}
When the impurity spins are put on opposite surfaces of the slab, they may communicate via electron states that are present in the bulk of the Weyl semimetal.  To examine this effect
numerically, we place the two spins on different surfaces of a WSM slab with various thicknesses and vary their separation along the $x$ axis (i.e., the direction in which the Fermi arcs states disperse). Results from these are illustrated in Figs.~\ref{fig:two} and ~\ref{fig:withN}.

\begin{center}
	\begin{table}[h]
		\begin{tabular}{ | c | c | c |}
			\hline
		 $N_z$ & $\alpha$ & $\beta$  \\ \hline
		9 & $\approx$ 3.28 & $\approx$ 2.36\\
			\hline
		17 & $\approx$ 3.33 & $\approx$ 2.39 \\
			\hline
		33 & $\approx$ 3.24 & $\approx$ 2.32 \\
			\hline			
		\end{tabular}~~~~~~
			\begin{tabular}{ | c | c |}
		\hline
		$R$ & $\gamma$  \\ \hline
		20 &  $\approx$ 4.96\\
		\hline
		40 &  $\approx$ 5.02 \\
		\hline
		60 & $\approx$ 5.06\\
		\hline			
	\end{tabular}
		\caption{Assuming the couplings $J_{ii}$ between the impurity spins on two opposite surfaces go as $\sim (R^2+L_Z^2)^{-\alpha/2}$, the best fitted value of $\alpha$ is quoted in the left-most column for the result of Fig.~\ref{fig:two}(b) for $R/a$ between 50 and 80. For fixed $L_z$, the same can be fitted with $R^{-\beta}$, which is shown in the middle column. For fixed $R$, the results from Fig.~\ref{fig:withN}(b) can be fitted (for $N_z$ between 13 and 21) with $L_z^{-\gamma}$, which is shown in the right-most column.}
	\end{table}
\end{center}

For the range of parameters we examined, the symmetry properties of the spin coupling matrix turn out to be the same as when the spins are situated in the bulk and are
separated along the $z$ direction (see Appendix A). Numerically, when the two spins are located at sites $(x,y,z) = (0,0,0)$ and $(R,0,L_z=N_za)$,  the coupling between them is surprisingly strong despite the fact that they reside on different surfaces, decaying as $\sim (R^2+L_z^2)^{-\alpha/2}$, where $\alpha$ is between 3 and 4, for fixed $L_z$ and increasing $R$, showing a rather slow decay of the RKKY coupling and possibility of mean-field magnetic ordering. Fitting the decay with $\sim R^{-\beta}$, where the thickness $L_z$ is fixed, the coupling decay in even slower manner, with $\beta $ between 2 and 3. Results from the tight-binding simulation and from the low-energy wavefunctions both support these results.

As a function of the thickness $L_z$, for fixed $R$ the couplings initially increase and after attaining maximum values decrease rapidly. 
Results for both the analytical and tight-binding approaches for varying $N_z$ 
are illustrated in Fig.~\ref{fig:withN}. For fixed $R$, as a function of the thickness the coupling decays as $\sim L_z^{-\gamma}$ with $\gamma\approx 5$. This is the same falloff as for RKKY coupling in the bulk of the WSM~\cite{RKKYWSMSur1}, suggesting
that for large enough $L_z$ the coupling between spins on opposite surfaces is eventually
dominated by the bulk states.
The clear differences among the parameters $\alpha,\beta,\gamma$ capture the essential physics of the WSM system with Fermi arcs, and have been listed in the Table II.

The maximum coupling as a function of the thickness can be qualitatively understood as follows. As the thickness increases, the Fermi arcs localize increasingly firmly on the surfaces, increasing the surface density of states near the Fermi energy. This leads to an increase in the coupling between the impurity spins on the surfaces and the conduction electrons,
which can mediate intersurface interactions effectively when $L_z$ is not too large.
On the other hand, as $L_z$ increases, the 
number of conduction electron states which are sensitive to both surfaces decreases,
resulting in weaker coupling between spins on opposite surfaces. With increasing thickness, the competition between these two mechanisms gives rise to a critical thickness for which the coupling between spins placed on opposite surfaces maximizes.
As in our earlier argument for impurities on the same surface, with $k_0$ the only relevant momentum scale we expect this thickness to scale as $\sim 1/k_0$. We explore this in Appendix B.  Again this critical thickness also defines a thin-film limit; the values of this critical thickness obtained from inter-surface coupling are of the same scale as those obtained from the intra-surface coupling.


\section{SUMMARY AND DISCUSSION}\label{sec:disc}
In this work we have examined RKKY interactions among impurity spins on the surfaces of Weyl semimetal (WSM) slabs, using both an approach in which the wavefunctions of the WSM electrons are found in an analytical form, and a more fully numerical recursive Green's function technique.  We find that Fermi arc surface states play an important role in the RKKY coupling, creating couplings that are stronger and more long-range than
is found for impurities well-inside the bulk of the system.  Surprisingly, even the
coupling between spins on opposite surfaces can be relatively strong.  As a function of
film thickness, we find that the RKKY couplings are non-monotonic, with maxima that
can define a ``thin-film'' limit, in which the effects of {\it both} surfaces are
in some sense maximal.  The relative strengths and signs of different components of the RKKY couplings $J_{ij}$ can be understood using a simple model in which only surface
states associated with the Fermi arcs are retained, and in which the Fermi arcs
are perfectly straight.

The importance of electron states with strong support on the surfaces can be examined by comparing results
for geometries with Fermi arcs to ones without them.  Fig. ~\ref{fig:xzsurface}
illustrates RKKY coupling for spins on the same and opposite surfaces which are perpendicular
to the direction of separation between the Weyl nodes in the bulk, for which surface states
are not present.  The generally smaller scale of the resulting couplings supports the
idea that the Fermi arc states play a large quantitative role in setting the coupling scale.

The results presented to this point have been for vanishing chemical potential $\mu$,
where the only extended Fermi surfaces are due to the Fermi arcs, and the Fermi energy
passes directly through the Weyl nodes in the bulk.  In general, when $\mu \ne 0$ and
the Fermi wavevector $k_F \ne 0$ in the bulk, one expects $2k_F$ oscillations in the RKKY
coupling.  Results for $\mu \ne 0$ are presented in Fig. ~\ref{fig:finitemu}, for which
the oscillations are apparent.  The envelopes within which these oscillations occur
behave rather similarly to the results for $\mu=0$.

When a system is of order or thinner than a critical thickness $\sim 1/k_0$, our results show that
a proper treatment of RKKY interactions requires one to retain states from the Fermi arcs
of {\it both} surfaces, even if the two spins reside on the same surface.  For real systems,
such as TaAs~\cite{taas}, the typical separation of Weyl nodes is rather small (of the order of $k_0\approx 0.1\pi/a$) and thus we expect the critical thickness to be of the order of several tens to a hundred lattice spacings.  Such thicknesses are quite reasonable for thin-film semiconductor systems.

We conclude with some speculations about the kind of magnetic order these RKKY interactions might induce in the low temperature state of spin impurities on the surfaces of a WSM thin film. At large distances, the strongest couplings we find are for $J_{yy} = J_{zz}<0$ within
a single surface, suggesting the system will form a planar ferromagnet in its ground state.
The non-vanishing $J_{yz}$ and $J_{zy}$ couplings if large enough could induce spiral order;
while at short distances these can be larger than the diagonal elements, at long distances
the latter are significantly larger.  Given the relatively slow spatial decay of the RKKY interaction, it seems likely that the system will favor ferromagnetism.  Furthermore, the
sign of coupling for impurities on different surfaces suggests that the magnetization
of the two surfaces will be parallel to one another in the groundstate.  In principle at
low temperature such magnetic order should be detectable.  Moreover, with this type of order
one expects a magnetic disordering transition at finite temperature in the Kosterlitz-Thouless universality class,
which might be detected in thermal measurements or via spin transport in the system.  Finally,
the importance of the Fermi arc states in supporting such magnetic order could be tested
by comparing the behavior of slabs in which the surfaces support them to ones in which they do not.
We leave the investigation of these questions to future work.

{\it Acknowledgements} -- HAF thanks the NSF for support through Grant Nos. DMR-1914451, DMR-1506263 and DMR-1506460, by the US-Israel Binational Science Foundation. AK thanks support from SERB (Gov. of India) through grant ECR/2018/001443 and BRNS (Gov. of India) support through grant 58/20/15/2019-BRNS.

\vspace{1cm}
\section*{Appendix}

\begin{figure*}[ht]
	\centering
	\includegraphics[width=0.95\textwidth]{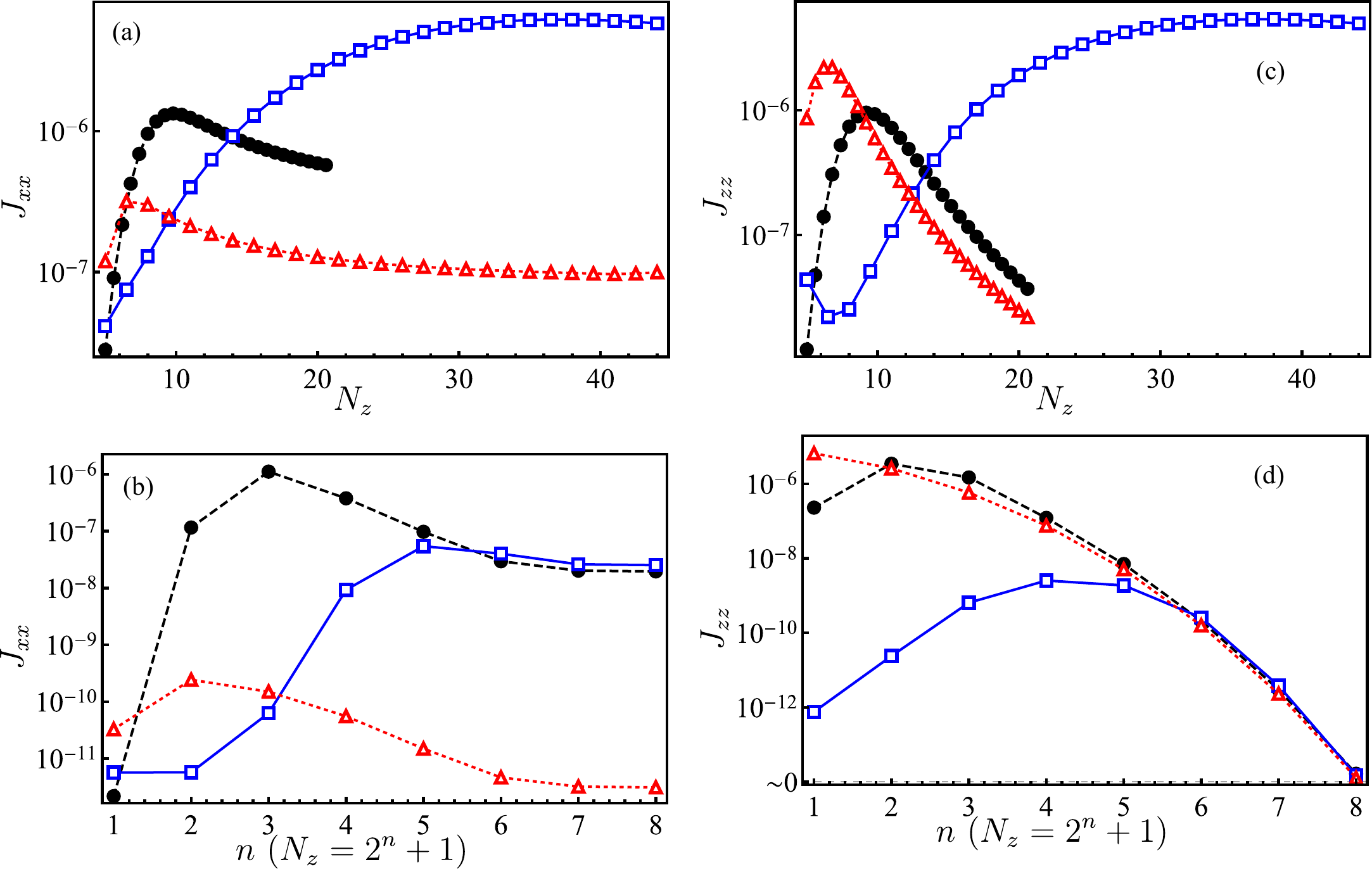}
	\caption{The critical thickness at which the RKKY coupling is maximum depends strongly on the separation of the Weyl nodes in momentum space, given by $k_0 = \cos^{-1}(m/\lambda)$. For three values of $m/\lambda =$ 0.2 (triangles), 0.5 (filled circles) and 0.95 (squares), (a) and (b) show the thickness dependence when the spins are on the same surface, while (c) and (d) show the thickness dependence when the spins are on opposite surfaces. For (a) and (c) the analytical solutions for wavefunctions of the low-energy Hamiltonian has been used.  For (b) and (d) the recursive Green's function method was used directly on the tight-binding model. The parameter values are the same as for Figs.~\ref{fig:top} and ~\ref{fig:two}. For (a) and (b), the two spins are at positions $(x,y,z)=(0,0,0)$ and $(R=40,0,0)$ and for (c) and (d) the two spins are at positions $(x,y,z)=(0,0,0)$ and $(R=40,0,L_z=N_za)$.}\label{fig:varyingm}
\end{figure*}

\section*{A. Symmetries among the susceptibility matrix elements}\label{app:symmetry}
In this appendix we briefly discuss the symmetries among the various coupling elements $J_{ij}$ for spins that are on the same or opposite surfaces, as well as in the bulk, based on the symmetry of the underlying Hamiltonian of the WSM.

First we consider the case when both spins, coupled to same orbital, are on the \textit{same surface} of a WSM slab of thickness $N_z$ with spatial separation $(x=R,0,0)$. The two Green's functions $G(\textbf{r}_{12},\omega)$ and $G(-\textbf{r}_{12},\omega)$ required to calculate the RKKY coupling between the spins for some arbitrary values of $x$ and $\omega$ have the structures
\begin{align}
&G(\textbf{r}_{12},\omega) = \left(\begin{array}{cccc}
s_0 & s_1 & s_2 & s_3\\
s_1& s_0 & s_4& s_2\\
 s_2& -s_3 & s_0 & s_1\\
-s_4 & s_2 & s_1 & s_0
\end{array}\right),\\
&G(-\textbf{r}_{12},\omega) = \left(\begin{array}{cccc}
s_0 & -s_1 & -s_2 & s_3\\
-s_1& s_0 & s_4& -s_2\\
-s_2& -s_3 & s_0 & -s_1\\
-s_4 & -s_2 & -s_1 & s_0
\end{array}\right),
\end{align}
where $s_{i}$ (with $i=0,1,2,3,4$) are complex numbers depending on $R$ and $\omega$. Using these in Eq.~\ref{eq:corr}, we obtain all possible nonzero components of the spin-spin correlation matrix to be
\begin{align}
\mathcal{J}_{zz}=\mathcal{J}_{yy},~\mathcal{J}_{xx}\neq 0~{\rm and}~\mathcal{J}_{zy}=-\mathcal{J}_{yz}.
\end{align}

Next, we consider the case when both spins, coupled to the same orbital, are on the \textit{opposite surfaces} of WSM slab of thickness $L_Z$ with lateral spatial separation $R$, the positions of the two spins are (0,0,0) and $(x=R,0,L_z)$. The two Green's functions $G(\textbf{r}_{12},\omega)$ and $G(-\textbf{r}_{12},\omega)$ required to calculate the RKKY coupling between the spins for arbitrary values of $R$ and $\omega$ now have the structure
\begin{align}
& G(\textbf{r}_{12},\omega) = \left(\begin{array}{cccc}
o_0 & o_2 & 0 & o_3\\
o_2& o_1 & -o_3& 0\\
0& -o_3 & o_0 & o_2\\
o_3 & 0& o_2 & o_1
\end{array}\right),\\
& G(-\textbf{r}_{12},\omega) = \left(\begin{array}{cccc}
o_1 & -o_2 & 0 & o_3\\
-o_2& o_0 & -o_3& 0\\
0& -o_3 & o_1 & -o_2\\
o_3 & 0& -o_2 & o_0
\end{array}\right),
\end{align}
with $o_{i}$ (with $i=0,1,2,3,4$) complex numbers depending on $R$, $L_z$, and $\omega$ ($o_2$ is two orders magnitude smaller than other elements). Using these in Eq.~\ref{eq:corr}, we find all possible nonzero components of spin-spin correlation matrix are related by
\begin{align}\label{eq:twosymm}
\mathcal{J}_{xx}&\approx\mathcal{J}_{yy},~\mathcal{J}_{zz}\neq0~,~\mathcal{J}_{xy}=-\mathcal{J}_{yx},\nonumber\\
\mathcal{J}_{zx}&=\mathcal{J}_{xz}~{\rm and}~\mathcal{J}_{zy}=-\mathcal{J}_{yz}.
\end{align}

Finally, we consider the case when the two spins, coupled to the same orbital, are deep in the bulk of a WSM and have spatial separation $(0,0,z=R)$. The two Green's functions $G(\textbf{r}_{12},\omega)$ and $G(-\textbf{r}_{12},\omega)$ required to calculate the RKKY coupling between the spins for some arbitrary values of $R$ and $\omega$ have the structure
\begin{align}
& G(\textbf{r}_{12},\omega) = \left(\begin{array}{cccc}
b_0 & 0 & 0 & b_2\\
0& b_1 & -b_2& 0\\
0& -b_2 & b_0 & 0\\
b_2 & 0 & 0 & b_1
\end{array}\right),\\
&G(-\textbf{r}_{12},\omega) = \left(\begin{array}{cccc}
b_1 & 0 & 0 & b_2\\
0& b_0 & -b_2& 0\\
0& -b_2 & b_1 & 0\\
b_2 & 0 & 0 & b_0
\end{array}\right),
\end{align}
where $b_{i}$, (with $i=0,1,2,3,4$) are complex numbers depending on $R$ and $\omega$. Using these in Eq.~\ref{eq:corr}, we obtain the components of the spin-spin correlation matrix as similar as those in Eqs.~(\ref{eq:twosymm}):
\begin{align}\label{eq:bulksymm}
\mathcal{J}_{xx}&=\mathcal{J}_{yy},~\mathcal{J}_{zz}\neq0~,~\mathcal{J}_{xy}=-\mathcal{J}_{yx},\nonumber\\
\mathcal{J}_{zx}&=\mathcal{J}_{xz}=0~{\rm and}~\mathcal{J}_{zy}=\mathcal{J}_{yz}=0.
\end{align}

\section*{B. Critical thickness and the separation of the Weyl nodes }\label{app:thickness}
Finally, we examine in more detail the $k_0$ dependence of the critical thicknesses at 
which the couplings are maximized, as discussed in the main text.
Fig.~\ref{fig:varyingm} illustrates numerical results for the thickness dependence of RKKY couplings for various values of the Weyl node separation $k_0$, for both the cases when the spins are on the same surface as well as when the spins are on the opposite surfaces. If one estimates the critical thickness $L_c$ at which the coupling attains its maximum,  one finds that $L_c k_0 \approx $ constant.


\begin{thebibliography}{99}
	\bibitem{WSMreview}
	For reviews see: N.~P.~Armitage, E.~J.~Mele, and A.~Vishwanath, Rev. Mod. Phys. \textbf{90}, 015001 (2018); Nature S. Jia, S.-Y. Xu, and M. Z. Hasan, Nature Materials \textbf{15}, 1140–1144 (2016); S. Rao, Journal of the Indian Institute of Science, \textbf{96}, 2 (2016).

	

	\bibitem{RKKY} M.~A.~Ruderman and C.~Kittel, Phys. Rev. {\bf 96}, 99 (1954); T.~Kasuya, Prog. Theor. Phys. {\bf 16}, 45 (1956); K. Yosida, Phys. Rev. {\bf 106}, 893 (1957).
	
	
	
	\bibitem{RKKYSO1}
	H.~Imamura, P.~Bruno, and Y.~Utsumi,
	Phys. Rev. B {\bf 69}, 121303(R) (2004).
	
	\bibitem{RKKYSO2}
	A.~Schulz, A.~De.~Martino, P.~Ingenhoven, and R.~Egger,
	Phys. Rev. B {\bf 79}, 205432 (2009).
	
	\bibitem{RKKYWSM1}
	H.-R.~Chang, J.~Zhou, S.-X.~Wang, W.-Y.~Shan, and Di Xiao,
	Phys. Rev. B {\bf 92}, 241103(R) (2015).
	
	\bibitem{RKKYWSM2}
	Yong Sun and Anmin Wang, J. Phys.: Condens. Matter \textbf{29}, 435306 (2017).
	
	\bibitem{RKKYWSM3}
	Mir Vahid Hosseini and Mehdi Askari,
	Phys. Rev. B \textbf{92}, 224435 (2015).
	
	\bibitem{RKKYQSH1}
	J.~Gao, W.~Chen, X.~C.~Xie, and F.-C.~Zhang,
	Phys. Rev. B {\bf 60}, 241302(R) (2009).
	
	\bibitem{RKKYQSH2}
	G.~Yang, C.-H.~Hsu, P.~Stano, J.~Klinovaja, and D.~Loss,
	Phys. Rev. B {\bf 93}, 075301 (2016).
	
	
	\bibitem{RKKYQSH1}
	J.~Gao, W.~Chen, X.~C.~Xie, and F.C.~Zhang,
	Phys. Rev. B {\bf 80}, 241302(R) (2009).
	
	\bibitem{RKKYTI}
	Q.~Liu, C.-X.~Liu, C.~Xu, X.-L Qi, and S.-C.~Zhang,
	\prl\ {\bf102}, 156603 (2009).
	
	\bibitem{RKKYQSH2}
	G.~Yang, C.-H.~Hsu, P.~Stano, J.~Klinovaja, and D.~Loss,
	Phys. Rev. B {\bf 93}, 075301 (2016).
	
	\bibitem{RKKY2DTI}
	C.-H.~Hsu, P.~Stano, J.~Klinovaja, and D.~Loss,
	\prb\ {\bf96}, 081405(R) (2017).
	
	\bibitem{RKKY2DTI2}
	C.-H.~Hsu, P.~Stano, J.~Klinovaja, and D.~Loss,
	\prb {\bf 97}, 125432 (2018).
	
		
	\bibitem{RKKYElectrical1}
	Y.-W.~Lee and Y.-L.~Lee,
	Phys. Rev. B {\bf 91}, 214431 (20015).
	
	
	
	%
	%
	\bibitem{Reja_2017}
	Sahinur Reja, H.A. Fertig, L. Brey, and Shixiong Zhang, Phys. Rev. B {\bf 96}, 201111 (2017).
	
	\bibitem{Reja_2019}
	Sahinur Reja, H.A. Fertig, and L. Brey, Phys. Rev. B {\bf 99}, 045427 (2019).
	
	\bibitem{SonuRKKY}
	Sonu Verma and Arijit Kundu,
	Phys. Rev. B \textbf{99}, 121409(R) (2019).
	
	
	\bibitem{RKKYWSMSur1}
	V. Kaladzhyan, A. A. Zyuzin, and P. Simon,
	Phys. Rev. B \textbf{99}, 165302 (2019).
	
	\bibitem{RKKYWSMSur2}
	Hou-Jian Duan, Shi-Han Zheng, Pei-Hao Fu, Rui-Qiang Wang, Jun-Feng Liu, Guang-Hui Wang, and Mou Yang,  New J. Phys. \textbf{20}, 103008 (2018).
	
	\bibitem{RKKYWSMSur3}
	Da Ma, Hua Chen, Haiwen Liu, and X. C. Xie,
	Phys. Rev. B \textbf{97}, 045148 (2018).
	
	
	\bibitem{WSMModel}
	M. M. Vazifeh and M. Franz, Phys. Rev. Lett. 111,
	027201 (2013).
	
	
	\bibitem{Keshav}
	Keshav Pareek and Arijit Kundu,
	arXiv:1812.05504 (unpublished).
	
	
	\bibitem{taas}
B. Q. Lv, H. M. Weng, B. B. Fu, X. P. Wang, H. Miao, J. Ma, P. Richard, X. C. Huang, L. X. Zhao, G. F. Chen, Z. Fang, X. Dai, T. Qian, and H. Ding,
Phys. Rev. X \textbf{5}, 031013 (2015).

	
	
	
\end{thebibliography}
\end{document}